\documentstyle[12pt,fleqn,epsfig]{article}
\begin{document}

\begin{center}
{\bf INSTITUT~F\"{U}R~KERNPHYSIK,~UNIVERSIT\"{A}T~FRANKFURT}\\
D - 60486 Frankfurt, August--Euler--Strasse 6, Germany
\end{center}

\hfill IKF--HENPG/3--98

\vspace{0.5cm}

\vspace{0.5cm}
\begin{center}
   {\Large \bf On $J/\psi$ Production in Nuclear Collisions}  
\end{center}

\vspace{1cm}

\begin{center}
Marek Ga\'zdzicki\footnote{E--mail: marek@ikf.uni--frankfurt.de}\\
\vspace{0.3cm}
Institut f\"ur Kernphysik, Universit\"at Frankfurt \\
August--Euler--Strasse 6, D - 60486 Frankfurt, Germany\\[0.8cm]
\end{center}

\begin{abstract}
\noindent
Data on $J/\psi$ production in inelastic proton--proton,
proton--nucleus and nucleus--nucleus interactions at 158
A$\cdot$GeV are analyzed and
it is shown that the ratio of mean multiplicities
of $J/\psi$ mesons and pions is the same for all these
collisions.
This observation is difficult to understand within current models
of $J/\psi$ production in nuclear collisions based on the assumption
of hard QCD production of charm quarks.
\end{abstract}

\newpage

\section{Introduction}

According to the factorization theorem 
of perturbative QCD \cite{Co:89}
inclusive cross section
of  a hard process should increase proportionally
to $A$ in p+A interactions and proportionaly to
$A^{2}$ in A+A collisions.
Models describing  $J/\psi$ 
production are built on the basis of this
prediction  (for review see \cite{Sa:95}).
They treat creation of a $c\overline{c}$
pairs 
as a hard process
and they further assume that the initial number of $J/\psi$
mesons is proportional to the number of
charm pairs. 
Therefore in the absence of medium effects the $J/\psi$
cross section is expected to  
increase as $A(A^{2})$
for p+A (A+A) collisions.

Experimental results on $J/\psi$ production
in p+A interactions contradict this naive expectation
showing an increase of the cross section proportional
to $A^{0.9}$.
This reduction of the $A$--dependence is usually 
explained as being predominantly due to
final state interactions of the $J/\psi$ meson (or its premeson state)
with nucleons \cite{Ge:92}.
However models based on this picture and parameters
fitted to the p+A data in general  overpredict
recent results on $J/\psi$ production in
central Pb+Pb collisions at 158 A$\cdot$GeV
(for review see  \cite{Ge:98}). 
This reduction of $J/\psi$ production
in the latter data  
is usually interpreted as due to interactions of $J/\psi$ (pre)mesons
with surrounding  high density matter (ultimately
the Quark Gluon Plasma)  
\cite{Sa:97}.

In high energy A+A collisions (from central S+S to central Pb+Pb) 
the multiplicity of pions and strange hadrons  increase 
proportionally to the number of colliding nucleons
(participant nucleons) \cite{Ga:97}.
These data and their interpretation in terms of
a statistical QGP model \cite{Ga:98}
suggested the question whether  a similar dependence may be
observed for charm and consequently for $J/\psi$ production.
A simple estimation of the centrality
dependence of the $J/\psi$ to pion ratio in Pb+Pb collisions
indicates that  this hypothesis may be in fact correct \cite{Ga:98}.

The aim of this paper is to review available experimental results
to obtain information concerning the $A$--dependence 
of $J/\psi$ yield.
In particular we  study the $A$--dependence of 
the $J/\psi$ to pion ratio using results on proton--proton
 and nucleus--nucleus  interactions
(Sections 2 and 3) and proton--nucleus interactions (Section 4).  
We summarize also results on the $A$--dependence of the open charm yield
in p+A interactions (Section 5).

\newpage

\section{ $J/\psi$ Multiplicity in p+p Interactions}

In p+p interactions
the $J/\psi$ cross section 
was measured  at five different 
collision energies, $\sqrt{s}$ = 6.8 GeV \cite{Ba:78},
8.7 GeV \cite{Co:81}, 19.4 GeV \cite{Ba:83}, 24.3 GeV
\cite{Na:75} and 52 GeV \cite{Mo:90}.
Most of the data are measured for $x_F > 0$ and they are not corrected
for the branching ratio to the measured decay channel.

The mean multiplicity of $J/\psi$ mesons
in full momentum space, $\langle J/\psi \rangle$,
is obtained in the following way.
The $x_F$ distribution of $J/\psi$ is assumed
to be symmetric with respect to reflection at $x_F$ = 0.
The most recent values of the branching ratios
$J/\psi \rightarrow \mu^+ + \mu^-$ ($B_{\mu\mu}$ = 0.0601)
and 
$J/\psi \rightarrow e^+ + e^-$ ($B_{ee}$ = 0.0602)
were used \cite{PDG}.
The cross sections of $J/\psi$ production were 
further divided by the
cross sections for inelastic p+p interactions
at the corresponding
collision energy. 
The latter cross sections were
calculated according to the parametrization
of the experimental data given in \cite{PDG}.
The resulting values of $\langle J/\psi \rangle$
are given in Table 1.

For  comparison with the data on nucleus--nucleus collisions
at the SPS 
the mean $J/\psi$ multiplicity for p+p
interactions at 158 GeV ($\sqrt{s}$ = 17.3 GeV) is needed.
In this energy range
the energy dependence of the integrated cross section for
$J/\psi$ production can be conveniently parametrized by
\cite{Sc:94}:
$$
\sigma^{J/\psi} = \sigma_0 (1- \frac {m_{J/\psi}} {\sqrt{s}} )^{a},
$$
where $a = 12$ and $\sigma_0$ are parameters fitted to the data,
$\sigma^{J/\psi}$ and $m_{J/\psi}$ are 
 $J/\psi$ cross section and mass, respectively.
This parametrization predicts a decrease of the $J/\psi$
yield  by about 25\% when going from $\sqrt{s}$ = 19.4 GeV
to $\sqrt{s}$ = 17.3 GeV.
Thus we can estimate $\langle J/\psi \rangle$ to be
$(2.9 \pm 0.5) \cdot 10^{-6}$ at $\sqrt{s}$ = 17.3 GeV using
the measured value of $\langle J/\psi \rangle$ = 
$(3.8 \pm 0.3) \cdot 10^{-6}$ at  $\sqrt{s}$ = 19.4 GeV 
(see Table I).
The result of the above  interpolation, shown by the open circle in Fig. 1,
agrees with the value
estimated  
in Ref. \cite{Sc:94} 
for p+p interactions at 150 GeV using 
data available
at this energy \cite{Ba:83} and an additional assumption 
concerning an unpublished ratio of  cross sections.

The mean multiplicity of negatively charged hadrons (more than 
90\% are $\pi^-$--mesons)
in nucleon--nucleon (N+N) interactions
at 158 GeV is   $\langle h^- \rangle$ =  $3.01 \pm 0.06$ 
\cite{Ga:95}.
This mean multiplicity was calculated as
$\langle h^- \rangle = (\langle h^- \rangle_{pp} + 2\langle h^- \rangle_{pn} +
\langle h^- \rangle_{nn})/4$,
where $\langle h^- \rangle_{pp}$, $\langle h^- \rangle_{pn}$ and
$\langle h^- \rangle_{nn}$ are mean multiplicities of negatively
charged hadrons for  p+p, p+n and n+n interactions at 158 A$\cdot$GeV,
respectively \cite{Ga:95}.

Taking the value of $\langle J/\psi \rangle$ calculated above
we obtain  
$\langle J/\psi \rangle/\langle h^- \rangle = (0.96 \pm 0.17)\cdot10^{-6}$
for N+N interactions at  $\sqrt{s}$ = 17.3 GeV.
This ratio is further used for the comparison with nucleus--nucleus data. 

\newpage

\section{$J/\psi$ Production in Nucleus--Nucleus
Collisions}

The production of $J/\psi$  in nucleus--nucleus collisions
was measured by the NA38 Collaboration  for
O+Cu, O+U and S+U interactions at 200 A$\cdot$GeV \cite{Ba:91} and
by the NA50 Collaboration
for Pb+Pb interactions at 158 A$\cdot$GeV
\cite{Ab:97, Ro:98}.
The procedure which allows to calculate the 
$\langle J/\psi \rangle/\langle h^- \rangle$ ratio from the
published data is described below using as an example
the Pb+Pb results.

The measured  $J/\psi$ cross section in minimum
bias Pb+Pb collisions is:
$$
B_{\mu\mu} \sigma^{J/\psi}_{acc} = 21.9 \pm 0.2 \pm 1.6~{\rm \mu b}.
$$
This cross section refers to the NA50 acceptance
$0< y_{cm} <1$ and $-0.5< cos\theta_{CS} <0.5$, where
$y_{cm}$ is the $J/\psi$ rapidity calculated in the c.m. system and
$\theta_{CS}$ is the Collins--Soper angle \cite{Co:77}.
In order to get an estimate of the total $J/\psi$
cross section we assume that the $J/\psi$ production
for $y_{cm} > 1$ can be neglected and that the distribution in
$cos\theta_{CS}$ is uniform \cite{Ba:83}.
This leads to a correction factor for the 
acceptance equal to 4.
Based on the h+p results at 200 GeV \cite{Ba:83} one can estimate 
that neglecting $J/\psi$ yield at $y_{cm} > 1$ may lead to an 
underestimation 
of the $J/\psi$ multiplicity  by less than 30\%.
A similar conclusion is reached when the $J/\psi$ rapidity distribution
in Pb+Pb collisions is assumed to be similar to  the
rapidity distribution of the $\phi$ mesons measured by the  NA49
Collaboration \cite{Fr:97}.
In addition, the cross section presented by  NA50
is corrected here for the branching ratio
$J/\psi \rightarrow \mu^+ + \mu^-$ ($B_{\mu\mu}$ = 0.0601) \cite{PDG}. 
The cross section for  $J/\psi$ production resulting from the above 
procedure is: 
$$
\sigma^{J/\psi} = 1.46 \pm 0.12 ~{\rm mb},
$$
where systematic uncertainty of our extrapolation procedure
is not included in the quoted error.
The $J/\psi$ multiplicity can be calculated as
$$
\langle J/\psi \rangle = \frac {\sigma^{J/\psi}  } 
{\sigma } = (2.07 \pm 0.17) \cdot 10^{-4},
$$
where $\sigma$ is 
the total cross section of inelastic Pb+Pb collisions calculated to be
7040 mb using a parametrization of the measured data given in Ref.
\cite{An:83}. 

The results of the NA35 and NA49 Collaborations \cite{Ga:97}
indicate that the ratio of $\langle h^- \rangle$ to the
mean number of participant nucleons, $\langle N_P \rangle$, is 
the same for
central S+S and Pb+Pb collisions 158 A$\cdot$GeV and equal to
$\langle h^- \rangle/\langle N_P \rangle = 1.93 \pm 0.14$.
We assume therefore  the same value of the ratio  for
inelastic Pb+Pb collisions.
Using the mean number of participant nucleons for the latter collisions
calculated within the Fritiof model \cite{Pi:92} 
($\langle N_P \rangle = 102$) we get $\langle h^- \rangle = 197  \pm 14$.
This leads to 
$\langle J/\psi \rangle/\langle h^- \rangle = (1.05 \pm 0.11)\cdot10^{-6}$  
for inelastic Pb+Pb collisions at 158 A$\cdot$GeV.

A similar procedure is used to calculate the
$\langle J/\psi \rangle/\langle h^- \rangle$ ratio for
O+Cu, O+U and S+U interactions.
There are however two difference. 
The $J/\psi$ cross sections for oxygen and sulphur 
induced reactions are measured at 200 A$\cdot$GeV
\cite{Ba:91}.
Therefore
for the comparison with the results at 158 A$\cdot$GeV
the measured values are scaled down by 25\% according to the
energy dependence of the $J/\psi$ multiplicity established for
p+p interactions (see Section 2).
Due to projectile--target asymmetry the $x_F$ distribution of $J/\psi$
is expected to be not symmetric with respect to reflection at 
$x_F = 0$. The correction for this effect is negelected.

The ratios obtained for nucleus--nucleus collisions
and the corresponding ratio for N+N interactions at the
158 A$\cdot$GeV are shown in Fig. 2 as a function of
$\langle N_P \rangle$. 
It is surprising that  
the  ratio $\langle J/\psi \rangle/\langle h^- \rangle$
is similar for nucleon--nucleon and nucleus--nucleus  interactions. 
One should however keep in mind 
that the ratios for nucleus--nucleus collisions may be underestimated
by  up to 30\%.
We repeat that this uncertainty
is due to limited accepetance of the
$J/\psi$ measurement for nucleus--nucleus collisions.
This systematic error can be reduced when the results
on the rapidity or $x_F$ distributions are published.

Finally we note that the ratio $\langle hard~~process \rangle/\langle h^- \rangle $
is expected to increase by a factor of about 3 when going from N+N
to Pb+Pb interactions, where $\langle hard~~process \rangle$ 
denotes here a mean
multiplicity of any process for which the cross section in A+A collisions
increases as $A^2$.

\newpage

\section{$J/\psi$ Production in p+A Interactions}

The inclusive cross section for $J/\psi$ production
in p+A interactions is measured in the region $x_F >$ 0 and it
is usually parametrized as \cite{Ge:98}:
$$
\sigma^{J/\psi} = \sigma_0(J/\psi) \cdot A^{\alpha(J/\psi)},
$$
where $\sigma_0(J/\psi)$  and $\alpha(J/\psi)$ are parameters
fitted to the experimental data.
A strong
dependence of  $\alpha(J/\psi)$  on $x_F$
was recently measured by
the E866 Collaboration \cite{To:98} at 800 GeV.
The $x_F$ distribution of $J/\psi$ decreases by a factor
of about 10 from $x_F$ = 0 to $x_F$ = 0.4 \cite{Sc:95}.
Thus the $A$--dependence of the integrated $J/\psi$ cross section
in the region $x_F > 0$ is dominated by the dependence
measured close to $x_F = 0$.
The values of $\alpha(J/\psi)$ obtained from
$x_F$ integrated data ($x_F \geq 0$) or from the data close to $x_F = 0$ range
from 0.89 to 0.94. 
The results were obtained by
various experiments \cite{Ba:83,Al:91,Le:95,Fl:97} 
in the collision energy range 200--800 GeV and they were
compiled in 
\cite{Ge:98}.
The  $\alpha(J/\psi)$ values are shown in Fig. 3 as a function of
$\sqrt{s}$ (filled circles).

In order to compare the $A$--dependence  
of $J/\psi$ and $h^-$ production the parameter $\alpha$
was fitted here to  data \cite{Ga:95} 
on the total multiplicity
of negatively charged hadrons
produced in p+A interactions at 200 GeV and 360 GeV.
In the fit the multiplicity of proton--nucleon (p+N)
interactions at the corresponding energy was included.
This multiplicity was calculated  as 
$\langle h^- \rangle =  (\langle h^- \rangle_{pp} +
\langle h^- \rangle_{pn})/2$ 
\cite{Ga:95}.
Finally the $\alpha$ parameter fitted to the multiplicity
data was added to the $\alpha$ parameter obtained by the
fit to the inelastic cross section results ($\alpha = 0.72 \pm 0.01$)
\cite{Ca:79}.
The obtained values of $\alpha(h^-)$
($\alpha(h^-)$ = 0.88 $\pm$ 0.01 at 200 GeV and
$\alpha(h^-)$ = 0.90 $\pm$ 0.02 at 360 GeV)
are shown in Fig. 4 (open circles).
The values of $\alpha(h^-)$ are similar to the values of
$\alpha(J/\psi)$.
There is no  evidence for any significant 
energy dependence both for $\alpha(h^-)$ and $\alpha(J/\psi)$.
Similar values of the $\alpha$ parameter for $h^-$ and
$J/\psi$ production imply that the ratio
$\langle J/\psi \rangle(x_f>0)/\langle h^- \rangle$ 
is approximately independent of $A$ for p+A interactions at
high energy.
We note that the difference in the $\alpha$ parameter of 0.02
(typical for the values shown in Fig. 3) results in 
10\% change in the multiplicity ratio between
p+N and p+Pb interactions.
This can be compared to about 70\% increase of the ratio
$\langle hard~~process \rangle/\langle h^- \rangle$
expected when going from p+N to p+Pb interactions,
where $\langle hard~~process \rangle$ denotes here  a mean
multiplicity of any process for which
the cross section in p+A interactions increases as $A$
($\alpha = 1$).

The measurements of $J/\psi$ production in the backward
hemisphere ($x_F < 0$) are poor.
However,  
the experimental data \cite{To:98}
seems to indicate
that $\alpha(J/\psi)$ for $x_F < 0$ is similar
to $\alpha(J/\psi)$ for  $x_F > 0$
(we note that this is not the case for pion production).
This suggests that our conclusion concerning the similar
$A$--dependence of $h^-$ and $J/\psi$ production,
based on the $J/\psi$ data from the forward hemisphere
only, may remain unchanged when the $J/\psi$ results in full
phase space become available. 

\newpage
\section{Conclusions}

The main result of this paper is that 
the ratio of mean multiplicities
of $J/\psi$ mesons and pions is similar for inelastic
proton--proton, proton--nucleus and nucleus--nucleus 
interactions at 158 A$\cdot$GeV.
In our opinion this experimental observation justifies  a question 
whether the generally
accepted picture of $J/\psi$ creation
based on the factorization theorem of the perturbative QCD
and subsequent suppression of the $J/\psi$ yield by the
interactions with the surrounding medium is valid.
In this picture the observed scaling behaviour of the
data,
$\langle J/\psi \rangle/\langle h^- \rangle
\approx const(A)$,
can be treated only as due to accidental
cancelation of several large effects.

It is obvious that the mechanism of $J/\psi$ production
can not be understood without  data on open charm creation.
Published data on $D$ and $\overline{D}$ production in p+A interactions 
are insufficient.
The results on the $A$--dependence are summarized in Fig. 4,
where $\alpha(D,\overline{D})$ is shown as a function
of $x_F$ for interactions at 400 GeV \cite{Du:84} and
800 GeV \cite{Le:94}.
It is clear that these data do not allow to distinguish
between $\alpha \approx 1$, as usually assumed 
for charm production on the basis of perturbative QCD,
and  $\alpha \approx 0.9$,
the value obtained for pion and $J/\psi$ production.
Data on open charm production in nucleus--nucleus collisions do not
exist.
It is therefore crucial for our understanding
of the mechanism of charm creation
and $J/\psi$ production 
to measure open charm yields
in nucleus--nucleus collisions.

\vspace{1cm}
\noindent
{\bf Acknowledgements}

I thank Reinhard Stock for triggering publication
of this work.
I also thank him 
Andrzej Bia{\l}as, Leonid Frankfurt, Mark I. Gorenstein,  
J\"org H\"ufner, Helmut Satz, Peter Seyboth and
Herbert Str\"obele for discussions
and comments.
Claudie Gerschel and Carlos Lourenco helped me to understand
data on $J/\psi$ production.

\newpage

{\bf Table 1}
\noindent
The  results on mean multiplicity of $J/\psi$
mesons produced in p+p interactions.

\vspace{0.5cm}

\begin{tabular}{|c|c|c|}
\hline
                 &         &                   \cr 
 $\sqrt{s}$ [GeV]    &~~~~~~~~
~ $ \langle J/\psi \rangle \cdot 10^6$
~~~~~~~~&~~~~
~ Reference
~~~~~
\cr
      &           &                 \cr 
\hline
\hline
   6.8    & 0.021 $\pm$ 0.006  & \cite{Ba:78}  \cr
   8.7    & 0.075 $\pm$ 0.037  & \cite{Co:81}  \cr
  19.4    & 3.8   $\pm$ 0.3    & \cite{Ba:83}  \cr
  24.3    & 4.6   $\pm$ 0.8    & \cite{Mo:90}  \cr
  52.0    & 19.7  $\pm$ 8.7    & \cite{Na:75}  \cr
\hline
\end{tabular}

\newpage

\begin{figure}[p]
\epsfig{file=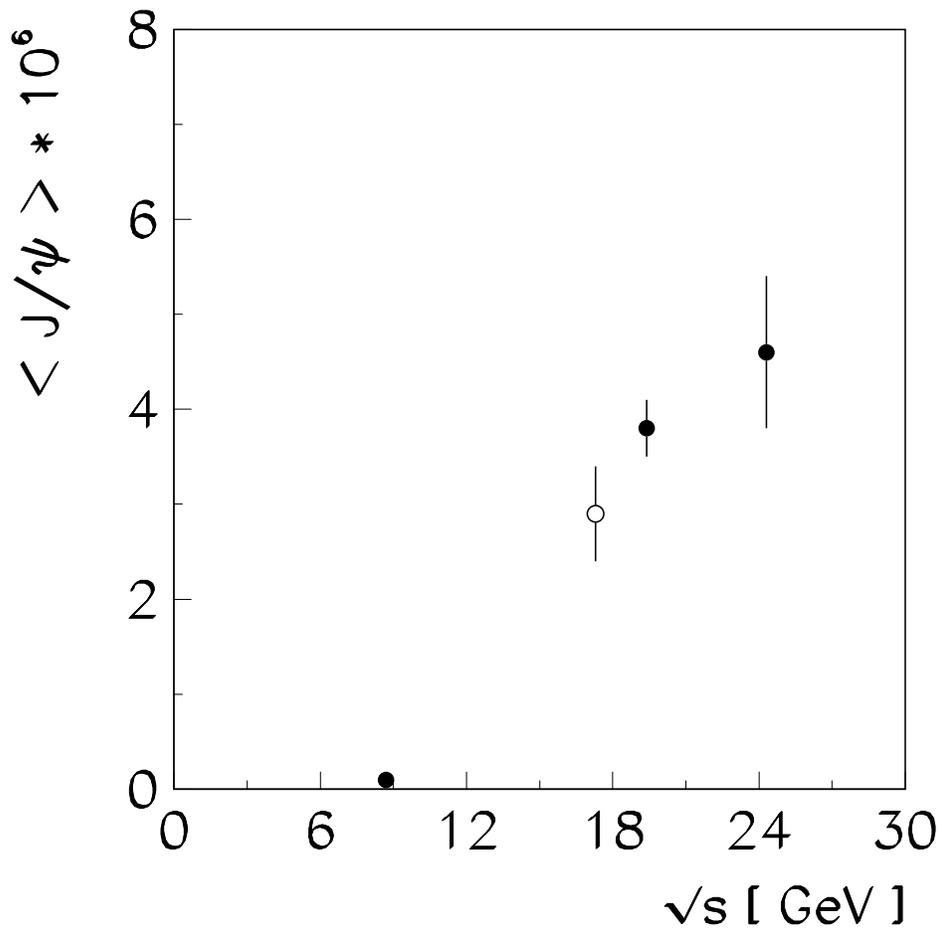,width=14cm}
\caption{ The multiplicity of $J/\psi$ mesons produced in
p+p interactions as a function of the collision energy.
The filled circles indicated measured data. 
The open circle shows the estimated multiplicity at $\sqrt{s} = 17.3$ GeV.
 }
\label{fig1}
\end{figure}

\newpage

\begin{figure}[p]
\epsfig{file=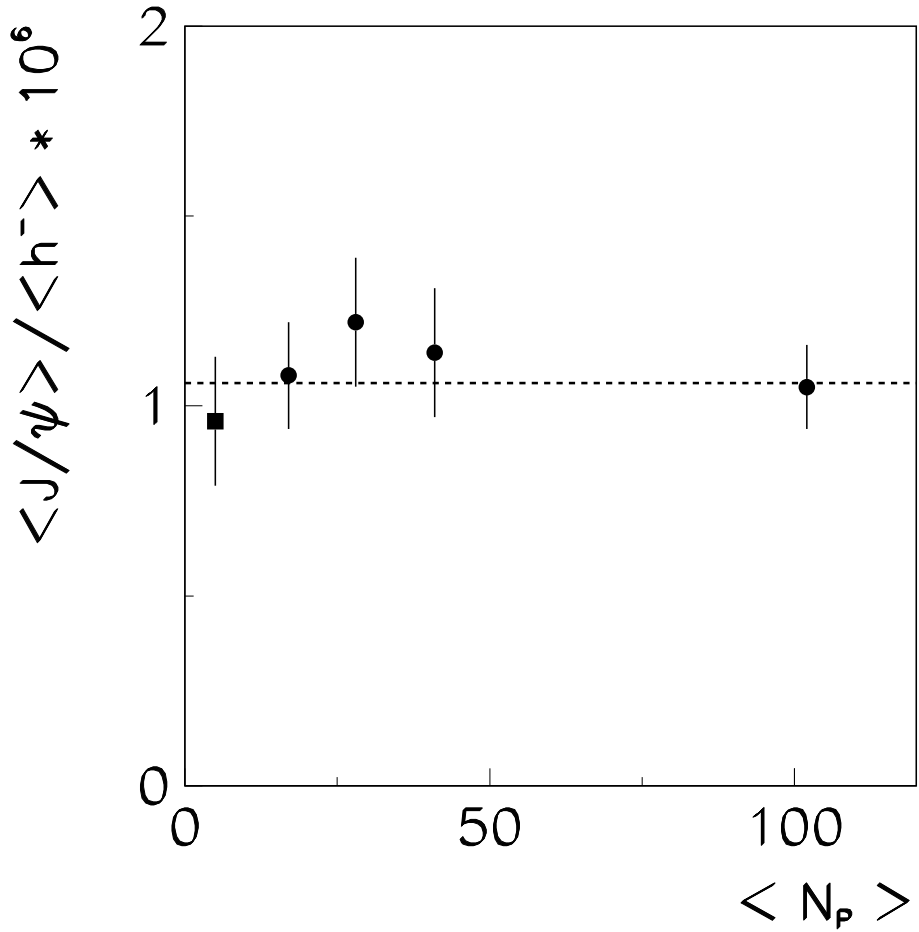,width=14cm}
\caption{
The ratio of the mean multiplicities of $J/\psi$ mesons
and negatively charged hadrons for inelastic nucleon--nucleon (square) and
inelastic O+Cu, O+U, S+U and Pb+Pb (circles) interactions at   
158 A$\cdot$GeV plotted as a function of the mean
number of participant nucleons.
For clarity the N+N point is shifted from
$\langle N_P \rangle = 2$ to $\langle N_P \rangle = 5$.
The dashed line indicates the mean value of the ratio.
}
\label{fig2}
\end{figure}

\newpage

\begin{figure}[p]
\epsfig{file=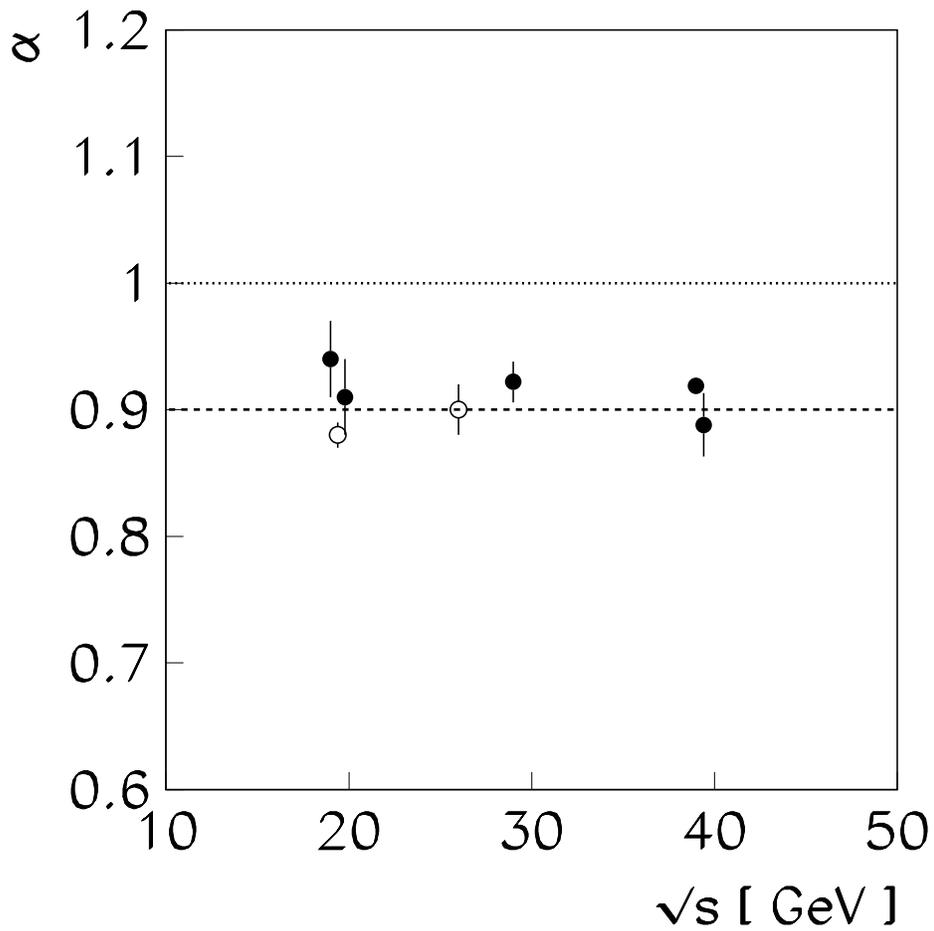,width=14cm}
\caption{ 
Comparison between $\alpha(J/\psi)$ (filled circles) 
and $\alpha(h^-)$ (open circles) for p+A interactions
in the energy range 200--800 GeV.
The dotted line shows the value $\alpha = 1$ 
characteristic for the  $A$--dependence of total
charm cross section obtained in models based on the perturbative QCD.
The dashed line indicates the value $\alpha = 0.9$
measured for pion production in full phase space.
}
\label{fig3}
\end{figure}

\newpage

\begin{figure}[p]
\epsfig{file=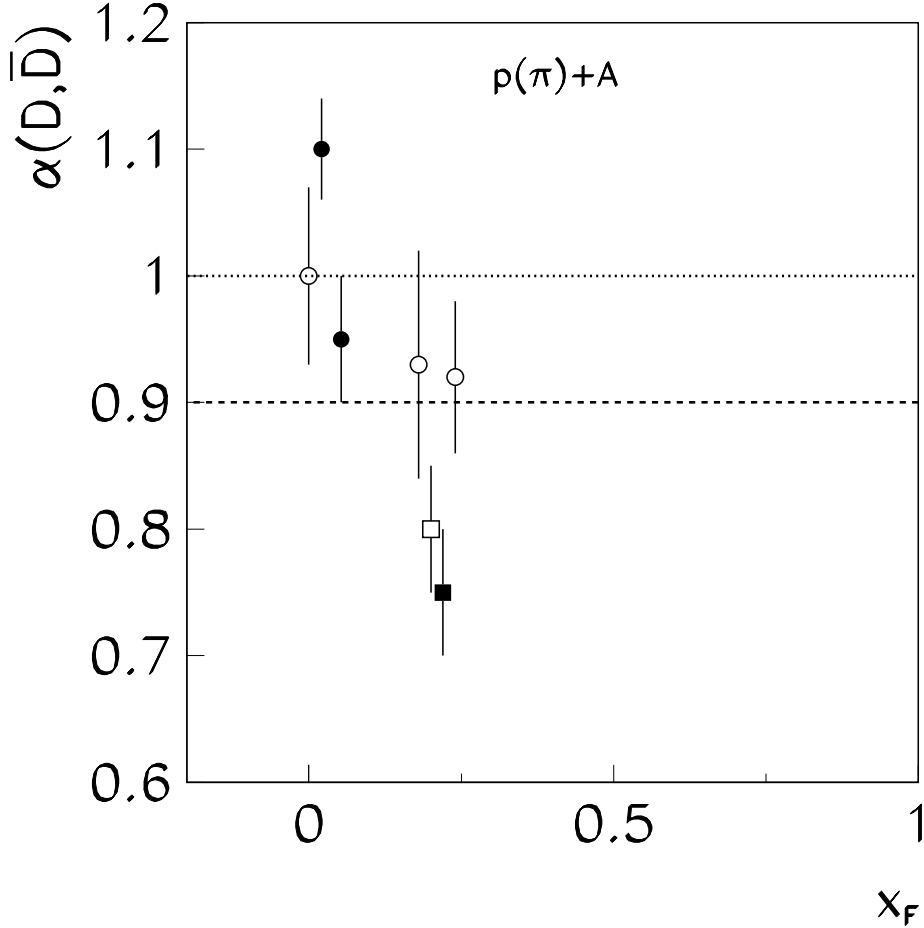,width=14cm}
\caption{ 
 Dependence of  $\alpha(D,\overline{D})$ on $x_F$
for p+A interactions \protect{\cite{Du:84,Le:94}} 
(filled symbols) and $\pi$+A interactions 
\protect\cite{Co:87,Al:92,Ad:92,Ad:97}
(open symbols) at 250--800 GeV. 
Circles indicate results obtained by 
reconstruction of $D$ and $\overline{D}$ decays.
Squares indicate data obtained by the analysis of prompt
single leptons or neutrinos.
The results for which the $\langle x_F \rangle$
is not given are plotted at the lower edge of the
acceptance region.
The dotted line shows the value $\alpha = 1$ 
characteristic for the  $A$--dependence of total
charm cross section obtained in models based on the perturbative QCD.
The dashed line indicates the value $\alpha = 0.9$
measured for pion production in full phase space.
}
\label{fig4}
\end{figure}

\end{document}